# Light Induced Aggregation of Specific Single Walled Carbon Nanotubes

*Madhusudana Gopannagari[†], and Harsh Chaturvedi[†] ***

[†]Department of Physics, Indian Institute of Science Education and Research, Pune-411008, INDIA

*Corresponding Author E-mail: hchaturv@iiserpune.ac.in

ABSTRACT: We report optically induced aggregation and consequent separation of specific diameter of pristine single walled carbon nanotubes (SWNT) from stable solution. Well dispersed solution of pristine SWNTs, without any surfactant or functionalization, show rapid aggregation by uniform exposure to UV, visible and NIR illumination. Optically induced aggregation linearly increases with consequent increase in the intensity of light. Aggregated SWNTs were separated from the dispersed supernatant and characterized using absorption and Raman spectroscopy. Separated SWNTs distinctly show enrichment of specific SWNTs under UV visible and NIR illumination.

Single walled carbon nanotubes (SWNT) are a material of active research for its diverse potential applications in electro-optics, plasmonics, biotechnology etc.[1-3] SWNTs are most commonly used as dispersed solution either for functionalization with other molecules, polymers, etc or for fabrication of devices, sensors.[4-6] Dispersion of SWNTs, as produced is a mixture of various diameters of metallic and semiconducting SWNTs.[7] Separation of SWNTs from solution is an important concern and various methods are being employed for developing efficient processes for separation of different SWNTs from solution.[8-10] As predicted by optical methods such as using photophoretic forces may play an important role for chiral and diameter specific separation of SWNTs from solution.[11] Photophoresis in aggregates of SWNTs have recently been reported by us.[12] Moreover, dispersion of nanoparticles like one dimensional SWNTs, also provides us with excellent opportunity to understand the effects of electro-optical forces on intermolecular interactions.[13] Optical absorption processes in one dimensional, pristine SWNTs have been well reported and characterized using spectroscopy.[14] Although optically induced aggregation especially in nanoparticles has theoretically been predicted and reasonably reported, very few experimental proofs of the same could be found.[15, 16] Optical induced aggregation in gold nanoparticles[17] and specifically in SWNTs functionalized with optically active supra-molecules has been reported by us.[18, 19] Stability of a solution depends on the dimensions, uniform surface charges and dielectric properties of the interacting particles and solution.[13] Under significant optical excitation, photophoretic forces in resonating particles are expected to affect the colloidal stability, leading to subsequent aggregation of the absorbing SWNTs. Here in we report, optically induced aggregation of selective pristine SWNTs from dispersed solution. Well-dispersed, stable solution of pure, pristine SWNTs in solution show enhanced rate of aggregation under optical illumination by simple UV, broadband visible and NIR lamps. Rate of aggregation shows dependence on the frequency and on the intensity of applied illumination. Significantly, the aggregated floc of SWNTs shows enrichment in selective nanotubes, corresponding to the frequency of the applied illumination. Hence, here-in we report, distinct separation of specific SWNTs by optically induced aggregation.

Well dispersed, stable solution of pristine SWNTs was prepared by ultrasonication of 0.6 mg of SWNT as purchased from (Nano Integris®-IsoNanotubes, 95% purity) in 100 mL of N,N-Dimethylformamide (DMF) (Sigma Aldrich®). Prepared dispersion was found stable for weeks. Care was taken to avoid water absorption and exposure to light. Fresh and well prepared disperse stable solution of pristine SWNT were exposed to UV lamp (125 W, 352 nm), broadband visible lamp (125 Watts) and NIR lamp (Fuzi, 150Watts) for different time periods, between 30-240 mins. Samples are also kept in dark for the same time periods, so as to find the normalized rate of aggregation for SWNTs in presence of light. Both the samples, ones kept in dark and ones exposed to light, were centrifuged @ 8000 rpm (Eppendorf® Mini Spin Centrifuge Machine) for 10 min to carefully separate the supernatant from the solution. Separated supernatant and flocs were analyzed by UV-Vis-NIR spectrophotometer (Perkin Elmer® Lambda 950 UV-Vis spectrometer) and Raman spectrometer (632 nm Laser).

Optically induced aggregation is observed in pristine SWNTs exposed to broadband visible, UV and NIR illumination. Rate of optically induced aggregation was determined using absorption spectroscopy as shown in **Figure 1(a-c)**. Concentration for each supernatant was normalized with reference to the stable concentration of the sample kept in dark. Concentration at 840nm was used to determine the rate of aggregation as the spectra here is relatively flat and free of van Hove singularities. Rate of aggregation is plotted with respect to time of exposure for each sample under different lamps. At the maximum exposure time interval of 240 minutes, about 20% and 40% of SWNT aggregates out under UV and broadband visible illumination respectively, however rapid aggregation is observed under NIR illumination with 65% SWNTs aggregating, as shown in **Figure 1 (d)**.

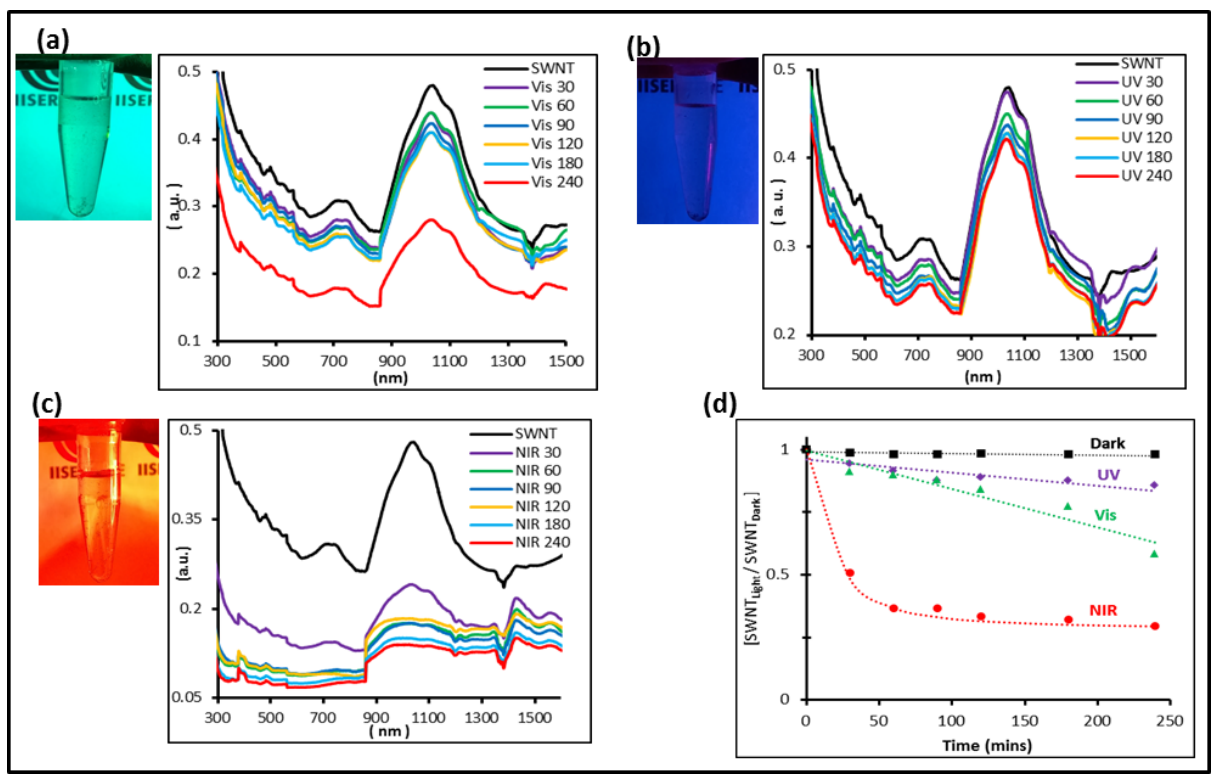

**Figure 1:** Light induced aggregation in broadband visible (a), UV (b) and NIR lamps (c) are shown. Absorption spectra of supernatant collected from pure SWNTs under UV, visible and NIR lamps show consistent decrease in concentration when exposed to varying duration from 30 mins to 240 mins (d) shows relative rate of optically induced aggregation in Pristine SWNTs under UV, visible and NIR illumination.

Further we study the dependence of this optically induced aggregation on the intensity of light as shown in the **Figure 2**. Well dispersed solution of pure SWNTs was uniformly exposed to varying optical intensity (25 W/m$^2$ – 200 W/m$^2$) from a broadband halogen lamp. Aggregated SWNTs were separated from the supernatant by centrifugation as mentioned above. SWNTs exposed to light show rapid aggregation and consistent decrease in concentration with increase in the intensity of light. Concentration of the supernatant separated from pristine SWNTs was normalized with the pristine SWNT kept in dark (control). **Figure 2(b)** show almost linear decrease in concentration of SWNTs with corresponding increase in the optical intensity.

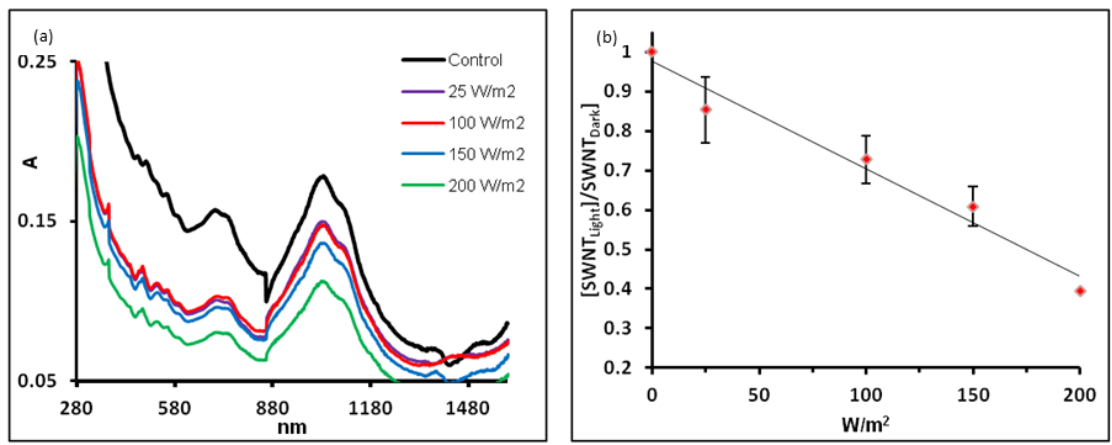

**Figure 2** (a) Absorption spectra of supernatant of pristine SWNTs under visible illumination with varying intensities. (b) SWNTs show enhanced rate of aggregation with increase in the optical intensity.

The absorption spectra of pure SWNTs show cumulative absorption by different diameters of nanotubes in solution. As shown in **Figure 1**, the absorption spectra of pure SWNTs shows background plasmonic resonance in UV (300-450 nm) due to metallic SWNTs and van hove singularities due to band-gap absorption in the visible and NIR by semiconducting SWNTs. **Figure 1(a-c)**, shows discernable changes in either the UV or in the visible-NIR frequency of the supernatant separated from the optically aggregated flocs, using different lamps. Supernatants separated from the solution exposed to NIR illumination for different times (**Figure 1c**), distinctly shows consistent changes and enrichment with specific SWNTs. Supernatant separated from the pure SWNTs, by exposing to visible illumination of different intensities also show significant changes in the visible and NIR frequency of the absorption spectra (**Figure 2**), indicating selective aggregation due to optical illumination. Moreover, **Figure 3(a)** show remarkable differences in the supernatant and aggregated floc, as separated from the pristine

SWNT solution, exposed to NIR illumination for 240 mins. Aggregated floc shows enrichment in SWNTs absorbing at 320 nm where-as supernatant shows remarkable loss of background resonance and enrichment with SWNTs absorbing at 380 nm. Significant changes are also observed in the NIR band (850-1500 nm) of the separated supernatant and floc. **Figure 3(b)**, shows absorption spectra of the separated aggregated floc under UV and NIR illumination for 240 min indicating distinctly different absorption features. Discernible changes are also observed in the RBM mode ( Figure 3 c) and G band (Figure 3 d) Raman spectra of the optically aggregated flocs separated from the pristine SWNT exposed to UV, visible and NIR illumination.

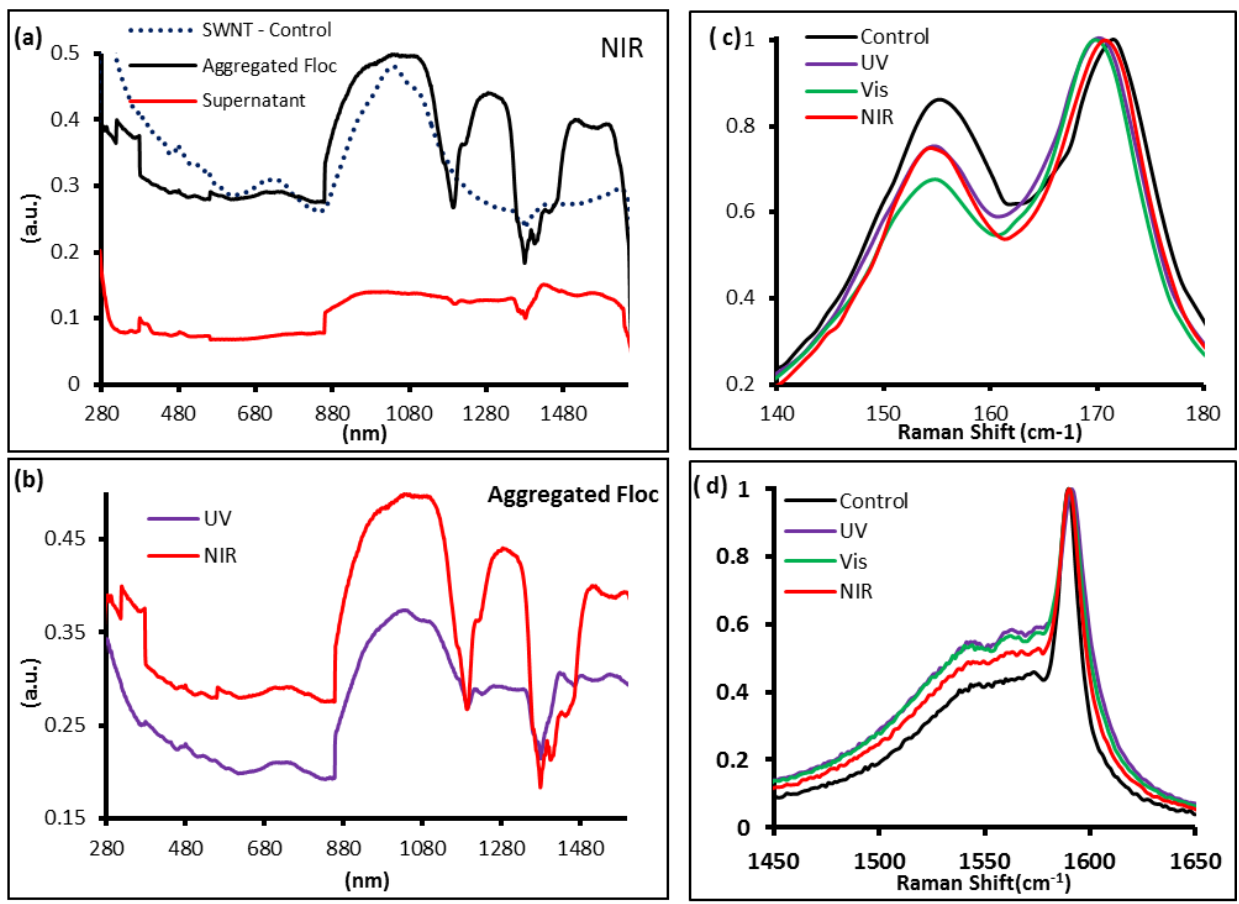

**Figure 3.** (a) Absorption spectra of the supernatant and aggregated floc as separated from the pristine SWNT (dotted line) after 240 mins exposure to NIR illumination. (b) comparing

absorption spectra of aggregated flocs separated after 240 mins from UV and NIR illumination. (c) Shows RBM raman spectra of separated aggregated flocs of SWNT, exposed to UV, visible and NIR spectra. (d) G band raman spectra of optically aggregated and separated flocs under different lamps.

RBM mode of pristine SWNTs depends on the diameter of the nanotube.[20] RBM of the aggregated SWNTs under different illuminations shows distinct enrichment of certain specific diameter of SWNTs which can be calculated using the relation $\omega_{RBM} = (\alpha_{RBM}/d) + \alpha_{bundle}$.[21] Where, $\alpha_{RBM}, \alpha_{bundle}$ are constants and d is the diameter of the SWNT corresponding to the RBM peak frequency($\omega_{RBM}$). As calculated from the expression, RBM of aggregates shows relative enrichment in following diameters 1.45, 1.46, 1.44 nm when exposed to UV, Visible or NIR illumination respectively. Consequent changes are also seen in SWNTs with larger diameters of 1.6 nm corresponding to the RBM frequency of ~ 1.55 cm$^{-1}$, Changes in the G band raman spectra of pristine SWNTs indicates significant changes in the relative concentration of metallic and semiconducting SWNTs.[22, 23] Figure 3 (d), shows significant changes in the normalized G$^-$ band, indicating increase in metallic SWNTs aggregating under UV and visible illumination as compared to NIR and un-separated SWNT solution. Experiments exploring the phenomenon for controlled separation of different diameter of SWNTs, along with detailed characterization and analysis are under progress. However, the results shown here do indicate aggregation of different diameter of SWNTs depending on the optical illumination. We believe, this phenomenon of optically induced aggregation in selective SWNTs is due to photophoretic and photo-thermal processes in SWNTs.[24] Induced photophoretic motion by similar lamps in aggregates of metallic or semiconducting SWNTs have recently been reported.[12] We believe similar photophoretic forces, along with photo-thermal processes are responsible for selective aggregation of pristine SWNTs. Thermal gradient on the surface of an absorbing particle in resonant optical frequency is given as $\nabla \hat{T} = \frac{1}{k_p} \hat{Q}_p$, where $k_p$ is internal heat conductivity of the particle and $\hat{Q}_p$ is volumetric thermal energy which relates to refractive index absorbed by the wavelength of incident radiation.[25] It is expected that, absorption and plasmonics resonance may lead to local temperature variations depending on the thermal conductivity of the absorbing SWNTs.

Even with inherent limitations due to various assumptions of surface charges, geometry and dielectrics, aggregation of SWNT dispersions are still widely explained using conventional DLVO theory.[26-28] DLVO theory describes total interaction energy between two particles in solution as a net result of attractive van der Waals (vdW) forces and repulsive electrostatic double-layer (EDL) interactions. This Interaction potential barrier, which defines the colloidal stability conventionally, depends on the dimensions, dielectric and surface charges at the particle-solution interface. For, $V_{tot} < 0$, van-der waals attractive term is greater than the repulsive EDL leading to colloidal instability and aggregation of the particles. Our experiments show significant decrease in $V_{tot}$ under optical illumination leading to enhanced rate of aggregation.

Total potential ($V_{tot}$) for two interacting cylinders in solution is approximated by the following expression.[15] $V_{tot} = \left( \frac{-A_H}{12\sqrt{2}D^{3/2}} + \frac{k^{1/2}}{\sqrt{2\pi}} Ze^{-kD} \right) \left[ \sqrt{\frac{R_1 R_2}{R_1 + R_2}} \right]$ Where, D is surface distance between two nanotubes with radius $R_1$ and $R_2$. $A_H$ is the Hamaker constant in the attractive Van der Waals contribution and Z is the interaction constant of the repulsive EDL term, $k^{-1}$ being the debye length. We believe this optically enhanced rate of aggregation is caused due to photophoretic forces causing local kinetics and non-uniformity of surface charges thus inducing colloidal instability.

Hence, in the present article we report a novel phenomenon of optically induced aggregation of selective SWNTs. Rapid aggregation was observed for pure SWNT solution exposed to optical illumination by simple lamps. Results show that the rate of aggregation primarily depends on the intensity of light with selective aggregation in specific SWNTs being caused due to frequency of illumination. Our results show that controlled optical excitation may be used for large scale efficient separation of particular diameter of SWNTs. Absorption and Raman spectra of the separated supernatant and aggregated floc especially in the NIR illumination convincingly show relative enrichment in specific SWNTs. Surely much research is indeed required both theoretically and experimentally to better understand and for developing efficient optical technology for separation of individual diameter of SWNTs. However, here we succinctly report the phenomenon of optically induced aggregation in selective SWNTs from pristine well dispersed solution, thereby causing separation of specific SWNTs.

ACKNOWLEDGMENTS

Authors are deeply indebted to Ramanujan fellowship (SR/S2/RJN-28/2009) and funding agencies DST (DST/TSG/PT/2012/66), Nanomission (SR/NM/NS-15/2012) for generous grants.